\documentclass[aip,jap,reprint]{revtex4-1}

\usepackage{graphicx}
\usepackage{bm}

\draft 

\begin{document}





\title{Relating Hysteresis and Electrochemistry in Graphene Field Effect Transistors} 








\author{Alina Veligura}
\email[a.veligura@rug.nl]{}
\author{Paul J. Zomer}
\author{Ivan J. Vera-Marun}
\author{Csaba J\'ozsa}
\author{Pavlo I. Gordiichuk}
\author{Bart J. van Wees}
\affiliation{Physics of Nanodevices, Zernike Institute for Advanced Materials, University of Groningen, The Netherlands}






\date{\today}

\begin{abstract}

Hysteresis and commonly observed p-doping of graphene based field
effect transistors (FET) was already discussed in reports over last
few years. However, the interpretation of experimental works
differs; and the mechanism behind the appearance of the hysteresis
and the role of charge transfer between graphene and its environment
are not clarified yet. We analyze the relation between
electrochemical and electronic properties of graphene FET in moist
environment extracted from the standard back gate dependence of the graphene resistance. We
argue that graphene based FET on a regular SiO$_2$ substrate
exhibits behavior that corresponds to electrochemically induced
hysteresis in ambient conditions, and can be caused by charge
trapping mechanism associated with sensitivity of graphene to the
local pH.

\end{abstract}


\maketitle 



\section{Introduction}

Graphene, as a single atom thick layer of carbon atoms, has already
showed potential for application in electronics and
biosensing\cite{Ratinac}. However graphene as a truly 2D system is
ultrasensitive \cite{Schedin2007} to the underlying substrate and
surface chemistry, which alters the charge transport properties of
pristine graphene. One of the main issues in graphene devices is a
hysteretic behavior of its resistance observed in ambient
conditions, when a gate voltage is swept back and forth. The
presence of hysteresis and commonly observed p-doping of graphene
based field effect transistors (FET) was already discussed in recent
reports\cite{Lafkioti,Wang2010,Sabri2009,Levesque}. The
interpretation of experimental works differs; and the mechanism
behind the appearance of hysteresis and the role of charge transfer
between graphene and its environment are not clarified yet.

In an ideal case of grounded graphene its charge neutrality point
(CNP) is located at zero back gate voltage. However, in ambient
conditions most of the graphene based FETs show initial p-doping
(CNP is positioned at positive Vg) and hysteresis. We point out that
these two effects can be related but do not necessarily have the
same nature. The doping of graphene can be caused either by the
adsorbates on top or underneath the graphene
surface\cite{Schedin2007,Lafkioti,Wang2010} or by the
electrochemical processes involving
graphene\cite{Sabri2009,Levesque,Sidorov2011}. Depending on the
nature of the dopant or the electrochemical environment, the initial
doping can be either p or n, which introduces a shift of the
graphene CNP to positive or negative gate voltages respectively. One
should keep in mind that even in the absence of a net doping the
dynamic response of the graphene resistance, namely hysteresis, can
be different.

There are two types of directions defined for hysteresis; positive
and negative \cite{Wang2010}. The positive direction of hysteresis
corresponds to the CNP shifting towards negative voltages while the
gate voltage is swept further into the negative regime. In case of
negative hysteresis the shift of the resistance with respect to the
gate voltage is in the opposite direction: the CNP shifts toward
more positive values while sweeping the gate into the negative
regime. Wang et al.\cite{Wang2010} proposed that negative and
positive hysteresis directions can be attributed to two competing
mechanisms: capacitive coupling and charge trapping from/to
graphene, respectively.

Capacitive coupling enhances the local electrical field near
graphene, inducing more charge carriers and causing a negative
direction of hysteresis. An example of a mechanism for capacitive
coupling is a dipole layer placed in between graphene and the back
gate. In moist air and without additional treatment of the silicon
oxide substrate (a common insulator for a GFET) this dipole layer
exists as adsorbed water molecules at room temperature
\cite{Moser,Lafkioti} or ordered ice at low
temperature\cite{Wang2010,Wehling}. The capacitive coupling
mechanism is also dominant in electrolyte-gating devices, via ions
in the electrical double layer \cite{Wang2010}. The positive
direction of hysteresis is caused by a charge trapping mechanism.
Accumulated charge in trap centers will start screening the electric
field of the back gate. One of the examples of trap centers are
surface states in between SiO$_2$ and graphene
\cite{Wang2010,Romero2008,Liao,Shin}. In case of graphene based FET
traps in bulk SiO$_2$ or SiO$_2$/Si interface were excluded in a
recent report by Lee et al.\cite{Lee2011}, who measured time scales
which were too fast for these types of trapped centers.

A separate charge transfer mechanism which was observed for the
hydrogenated surface of diamond \cite{Chakrapani}, carbon nanotubes
\cite{Aguirre} and graphene based FETs
\cite{Sabri2009,Levesque,Sidorov2011}, is the dissociation of
adsorbed water and oxygen on the carbon surface. Since water in
equilibrium with air is slightly acidic (pH=6), the electrochemical
potential of the carbon surface is higher than that of the solution,
resulting in electron transfer from graphene. Therefore, a graphene
FET possesses a net p-doping in moist air. The electron transfer is
mediated by oxygen solvated in water and can occur in opposite
direction with increasing pH. This redox can therefore influence the
dynamic response of graphene devices under an applied back gate and
cause a positive hysteresis.

A recent report by Fu et al.\cite{Fu} opened the discussion whether
graphene pH sensitivity is caused by charge transfer directly
between graphene and the solution \cite{Ang2008,Ohno,Heller} or if
the sensitivity is mediated by a layer on top or next to graphene
(either oxide or polymer residue). This layer can provide terminal
hydroxyl groups which can be protonized or deprotonized depending on
the proton concentration in the solution (pH), yielding a bound
surface charge layer, which can electrostatically induce carriers in
graphene. Recently it was reported that application of a gate potential
can lead to a local change of pH in a thin water film next to an
oxide substrate \cite{Veenhuis}. We argue that a combination of
these two effects can result in a positive hysteresis in graphene,
where the residues act as mediators for charge trapping actuated by
pH changes induced via gate electrical field. We emphasize that both
cases, independent whether the charge trapping is direct or mediated
by residues, would lead to the same direction in hysteresis and will
be undistinguishable in transport experiments. Though replacement of
the silicon oxide with either a hydrophobic \cite{Lafkioti,Shin} or
an oxygen free \cite{Sabri2009} substrate did show suppression of
both initial p-doping and hysteretic behavior, none of the reports
link the chemical redox to the direction of hysteresis.

In this work we analyze the relation between electrochemical and
electronic properties of graphene FET in moist environment. We argue
that graphene based FET on a regular SiO$_2$ substrate exhibits
behavior that corresponds to electrochemically induced hysteresis in
ambient conditions, caused by charge trapping mechanisms associated
with the sensitivity of graphene to the local pH.

\section{Methods}
Samples were obtained by mechanical exfoliation of graphite (Highly
Ordered Pyrolytic Graphite or Kish) on an oxidized n$^+$-doped
silicon substrate (300 or 500 nm thick oxide layer), which functions
as a back gate. The SiO$_2$ wafers are commercially available from
Silicon Quest International, where the oxide is prepared by dry
oxidation. Single layer graphene flakes were chosen based on their
optical contrast and thickness measured by atomic force microscopy.
A small number of samples were inspected with Raman spectroscopy to
verify the number of layers. Ti/Au (5/40 nm thick) electrodes were
prepared using standard electron beam lithography and lift off
techniques. For electrical measurements samples are placed in a
vacuum can with base pressure of $~5\cdot 10^{-6}$ mbar, using a
standard low frequency AC lock-in technique with an excitation
current of 100 nA. The carrier density in graphene is varied by
applying DC voltage (Vg) between the back gate electrode and the
graphene flake, as depicted in Fig.~\ref{fig:Fig1}(a). The charge
carrier mobilities ($\mu$) ranged from 2.500 up to 5.000 cm$^2$/Vs
at a charge carrier density of $n=2\cdot 10^{11}cm^{-2}$.

The sensor properties of the devices were studied in the following
way. First, we pumped down the sample can (95 cm$^3$ in volume) to
the base pressure. Then a valve connecting the can to a volume,
containing liquid water and filled with saturated vapor (H$_2$O or
D$_2$O at ~32 mbar saturation pressure) at 25 C$^o$, was kept open
for 1~s (short exposure to the vapor). After measurement, the valve
to the sample was fully opened, connecting the sample volume to the
water container (flooding with water vapor). In case of ethanol
vapor exposure the procedure was kept the same, but the partial
pressure of ethanol in the liquid cavity was 78 mbar. The purity of
heavy water and ethanol was 99.9\%.
 A graphene based FET on a hydrophobic substrate was also
prepared by exposure of SiO$_2$ to hexamethyldisilazane (HMDS) vapor
prior to graphene deposition. HMDS forms a self assembled monolayer
which protects graphene from the influence of dangling bonds in
silicon dioxide and prevents adsorbtion of water molecules in the
vicinity of graphene.

\section{Results and discussion}

\begin{figure}
\includegraphics[width=8.5cm]{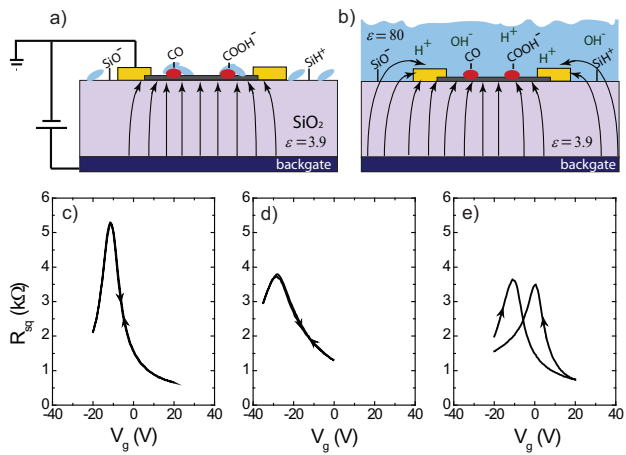}
\caption{\label{fig:Fig1}(Color online) a) Scheme of a graphene
based device with a discontinuous layer of adsorbed water in case
of a short exposure to H$_2$O vapor. Dangling bonds in SiO$_2$,
lithographic polymer remains (in red) on the graphene surface and electric
field lines between graphene and the back gate are schematically
drawn; b) A continuous thin layer of water on the graphene surface in case of
flooding the sample with water vapor; c) Graphene resistance versus
gate voltage after annealing (initial state); d) After a short
exposure to water vapor; e) Positive hysteresis developed
after further flooding with water vapor.}

\end{figure}

In ambient conditions the devices appear to be p-doped, with a
pronounced positive hysteresis in the dependence of resistivity
versus gate voltage (not shown). To remove adsorbates from the
graphene surface we perform global annealing of the device in vacuum
at 130$^o$C for 1,5 hrs. After annealing, the gate dependence does
not show hysteresis and becomes symmetric around the CNP
(Fig.~\ref{fig:Fig1}(c)), which is located at a negative gate
voltage (-11 V), indicating electron doping. Similar shifts towards
negative gate voltages were observed by Romero et
al.\cite{Romero2008} and associated with SiO$_2$ surface states. We
will call this position of the charge neutrality point the initial
position (after annealing). Short exposure to water does not cause
hysteresis, but reduces $\mu$ by 25 \% compared to the initial state
and can be attributed to the increase of a number of the
scatter centers for charge carriers\cite{Schedin2007}
(Fig.~\ref{fig:Fig1}(d)). Since graphene is hydrophobic, we assume
that during the short exposure adsorbates only occasionally
agglomerate on the graphene surface in the vicinity of polymer
leftovers which are unavoidably present after the lithography step
(Fig.~\ref{fig:Fig1}(a)).

Flooding the sample chamber with H$_2$O vapor assures full coverage
of the previously annealed SiO$_2$ and graphene surface with a thin
film of water (~3 nm thick \cite{Kim}), similar to ambient
conditions. After flooding we observe both electron-hole asymmetry
and a highly hysteretic behavior of the graphene device, where the
CNP for trace and retrace are situated at Vg of opposite signs
(Fig.~\ref{fig:Fig1}(e)). Moreover, a decrease of the scanning rate
in gate voltage sweeps (V/s) leads to more pronounced hysteresis
with the spacing between trace and retrace maxima increasing from
6.5 V at 1 V/s up to 23,5 V for 0,1~V/s. The cycle of annealing and
water exposure was repeated a few times showing reproducible
results. The positive direction of hysteresis indicates charge
trapping mechanism, while electron -hole asymmetry can be explained
in two ways: real asymmetry due to doping of graphene under the
contacts\cite{Huard} or an artifact of charging and discharging
graphene due to the hysteresis. Since we do not observe asymmetry in
the initial curve, the latter situation will be assumed in further
discussions.

Next, we present a novel analysis of hysteretic back gate voltage
sweeps from the point of view of time dependent shifts in CNP. These
shifts represent a change in carrier density within a certain time,
equivalent to a current. We estimate this current corresponding
either to the charge flow in or out of graphene, or induced charge,
in the following way. Charge current is extracted by comparing the
non hysteretic Dirac curve of graphene, which is shortly exposed to
water vapor, to the curves after the sample is flooded, measured at
different scan rates: 0,5; 0,25 and 0,1 V/s. The exact procedure is
shown in Fig.~\ref{fig:Hysteresis}a), b). For each scan rate the
gate voltage axis was divided into fixed regions $\Delta V_{fixed}$.
A change in voltage  $\Delta V_{fixed}$ induces a change in the
carrier density and resistance  $\Delta R$ accordingly. Due to the charge
trapping mechanism induced by water, the same $\Delta R$ will require a
different value of gate voltage  $\Delta V_i$ in case of the non-hysteretic
curve. The difference between $\Delta V_{fixed}$ and  $\Delta V_i$ will be
proportional to the amount of additionally induced or transferred
charge in graphene. The charge current (A/$\mu$m$^2$) in graphene
can then be calculated as:

\begin{figure}
\includegraphics[width=8.5cm]{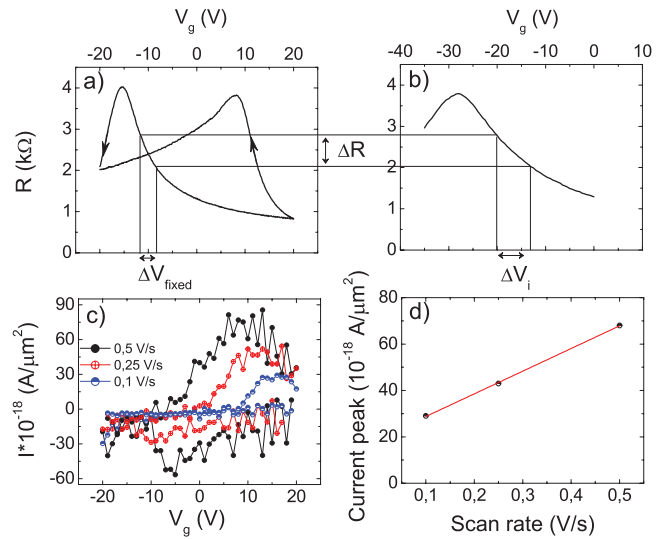}
\caption{\label{fig:Hysteresis} (Color online) Calculation of the
charge current in graphene. a) Gate voltage dependence of graphene
resistance "flooded" with water vapor and measured at a rate of 0,1
V/s. The curve is divided into parts with a fixed step in gate
voltage  $\Delta V_{fixed}$, corresponding to the change in resistance
$\Delta R$; b) Gate dependence of graphene resistance shortly exposed to H$_2$O
vapor. Due to the charge transfer now the same change $\Delta$R requires different value of
applied voltage  $\Delta V_i$ ; c) Calculated charge current versus gate
voltage for three different scan rates: 0,5; 0,25 and 0,1 V/s; d)
Linear scaling of the peak, at positive gate voltage shown in c), with the scan rate.
}
\end{figure}

\begin{equation}
I_{i} = \frac{e\alpha(\Delta V_i-\Delta V_{fixed})} {\Delta
V_{fixed}/\beta} \,
\end{equation}
where $e$ is the elementary charge,  $\alpha=2\cdot
10^{14}m^{-2}V^{-2}$ with $e\cdot\alpha$ the charge capacitance per unit area for 500 nm
SiO$_2$, and $\beta$ is the scan rate of the gate sweep (V/s).

The calculated charge current curves (Fig.~\ref{fig:Hysteresis}(c))
resemble the electrovoltaic characteristics of graphene based
electrochemical cells with controlled pH \cite{Fu}. A graphene based
device on a SiO$_2$ substrate can act as a working electrode in the
thin layer of water covering the hydrophilic oxide surface. Thus we
can consider graphene based devices as electrochemical cells.
Moreover, the height of the observed peaks scales linearly with the
scan rate of the applied gate voltage (Fig.~\ref{fig:Hysteresis}(d))
which, for an electrochemical cell, suggests that these peaks
originate from a non-Faradaic or non-diffusion limited process
involving the adsorbed ions on the graphene surface\cite{Ang2008}.
We performed the same sequence of experiments with graphene devices
on HMDS primed SiO$_2$. In contrast to graphene on hydrophilic
SiO$_2$ we observe neither hysteresis nor any changes in the
graphene resistance under water vapor exposure.\\

From the fact that the initial curve (after annealing) has no
hysteresis we can exclude charge trapping in the surface states of
SiO$_2$. Comparing to a local current annealing procedure
\cite{Wang2010}, here we globally annealed the sample which assures
desorption of H$_2$O molecules from the whole SiO$_2$ surface and
prohibits their diffusion back to the graphene surface. The
hysteresis appears only when the amount of water in the system is
high enough to form a continuous layer. The linear scaling of
extracted height of current peaks with scan rate indicates the
reversible charging of an ionic layer at the graphene surface
(electrode) by an applied gate voltage. The absence of hysteresis of
the graphene resistance when HMDS is used supports the idea that the
trapping mechanism happens by the presence of a water layer on the
SiO$_2$ surface. The dielectric constant of water is
$\varepsilon_{H2O}=80$ much higher than $\varepsilon_{oxide}=3.9$.
Therefore the electrical field lines in the device deviate from
plane capacitor and can be present in the water layer
(Fig.~\ref{fig:Fig1}(b)). The strong electrical field across the
water layer can either cause dissociation of water molecules
\cite{Teoh} or proton release/uptake by terminal OH$^-$ groups at
the oxide surface, as previously described\cite{Fu,Veenhuis}. Both
these mechanisms lead to a local pH change in the graphene vicinity.
Depending on the pH, the dangling bonds of the oxide or polymer
remains on graphene will change their charge state, inducing an
opposite charge in graphene\cite{Fu,Teoh}. At the present state we
can not pinpoint the exact identity of the ionic species causing the
change of environment around the graphene. A possible
electrochemical reaction on the unprotected Au electrodes is not
relevant as this was ruled out by Wang at al\cite{Wang2010}. where
both samples with protected and unprotected gold contacts showed the
same type of hysteresis.

\begin{figure}
\includegraphics[width=8.5cm]{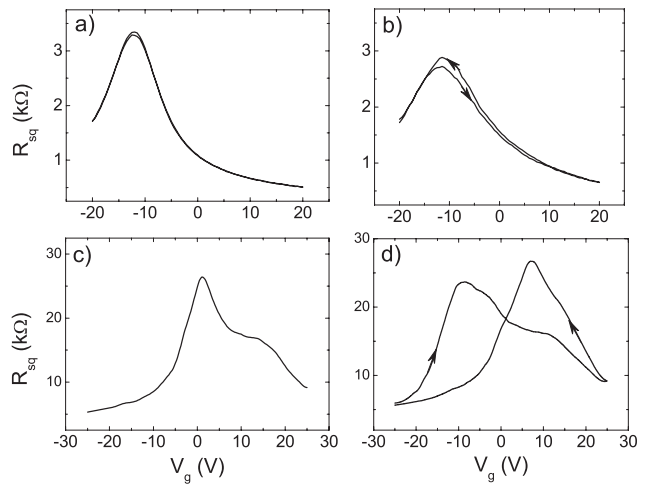}
\caption{\label{fig:Ethanol} Changes in graphene
resistance versus gate voltage under exposure to ethanol and D$_2$O vapors. a) The initial
state; b) After further flooding with ethanol vapor; c)  The initial
state (another sample); d) Previously mentioned sample after further flooding with D$_2$O vapor.
}
\end{figure}

Since the dipole nature of water molecules is often discussed in
relation to the hysteresis observed in graphene devices
\cite{Wang2010,Lafkioti,Wehling}, we decided to study the response
of graphene resistance to ethanol vapors. A pure neutral ethanol
solution has at least 100 times less concentration of H$^+$
 and OH$^-$ ions than pure water\cite{Hansen}. However
the dipole moment of an ethanol molecule $\vec{p}_e=1.68 D$ is
comparable to that of water $\vec{p}_w=1.85 D$ \cite{Hansen}, which
makes it possible to separate the electrochemical from electrostatic
influences on the charge carrier density in graphene. In
Fig.~\ref{fig:Ethanol}(a,b) the changes in graphene resistivity
under ethanol vapor exposure are presented. Except for the reduction
of charge carrier mobility by ~25 \% (comparable to water exposure)
neither considerable hysteresis nor doping were observed.

We also performed similar experiments using D$_2$O vapor with
another set of samples. Chemically, D$_2$O molecules behave similar
to H$_2$O. However, D$^+$ ions are two times heavier than H$^+$,
whereas  the relative increase in mass of OD$^-$ ions compared to
OH$^-$ is negligible. If the electrochemical process on graphene
surface is proton diffusion limited, one expects to observe a
different behavior of the hysteresis at various scan rates.
Experimentally we do not observe any significant difference in
graphene's response between H$_2$O and D$_2$O. Heavy water exposure
causes doping and direction of the hysteresis comparable to normal
water values (Fig.~\ref{fig:Ethanol}(c,d)).

Our experiment with ethanol vapor supports the idea that the
polarity of molecules adsorbed in the graphene vicinity does not
influence the dynamic response of graphene resistance to a gate
voltage. We suggest that the main reason of the observed hysteresis
in ambient conditions is the electrochemical activity of water
molecules in the graphene environment.









%




%











%






\section{Conclusions}

In conclusion, we have shown that the commonly observed positive
hysteresis in graphene FETs can be derived from the electrochemical
activity of water adsorbates on the SiO$_2$ substrate. In a moist
environment a standard graphene FET can act as an effective
electrochemical-cell with graphene being an electrode in the thin
layer of water. Therefore the application of the back gate voltage
may lead to local changes of pH which in turn affect the carrier
density in graphene. From this point of view we suggest that, next
to contact doping effect, the observed electron-hole asymmetry in
graphene resistance appears as an artifact of the hysteresis caused
by charge trapping. Conducted experiments with ethanol vapor and
heavy water did not show a relation between the hysteresis and
neither dipole moment nor mass of adsorbed molecules, supporting the
idea of electrochemical activity of water as a key element in the
dynamic response to gate voltage sweeping. These findings give a
further insight to graphene-related electrochemistry outside an
ideal electrochemical cell and open perspectives for the application
of a graphene FET as a memory element.

\begin{acknowledgments}
We would like to thank Bernard Wolfs, Siemon Bakker, and Johan G.
Holstein for technical assistance and Daniele Fausti for measuring
Raman spectra. This work is part of the research program of the
Foundation for Fundamental Research on Matter (FOM) and is supported
by NanoNed, NWO, and the Zernike Institute for Advanced Materials.
\end{acknowledgments}


\bibliography{}

\begin{thebibliography}{24}%
\makeatletter
\providecommand \@ifxundefined [1]{%
 \@ifx{#1\undefined}
}%
\providecommand \@ifnum [1]{%
 \ifnum #1\expandafter \@firstoftwo
 \else \expandafter \@secondoftwo
 \fi
}%
\providecommand \@ifx [1]{%
 \ifx #1\expandafter \@firstoftwo
 \else \expandafter \@secondoftwo
 \fi
}%
\providecommand \natexlab [1]{#1}%
\providecommand \enquote  [1]{``#1''}%
\providecommand \bibnamefont  [1]{#1}%
\providecommand \bibfnamefont [1]{#1}%
\providecommand \citenamefont [1]{#1}%
\providecommand \href@noop [0]{\@secondoftwo}%
\providecommand \href [0]{\begingroup \@sanitize@url \@href}%
\providecommand \@href[1]{\@@startlink{#1}\@@href}%
\providecommand \@@href[1]{\endgroup#1\@@endlink}%
\providecommand \@sanitize@url [0]{\catcode `\\12\catcode
`\$12\catcode
  `\&12\catcode `\#12\catcode `\^12\catcode `\_12\catcode `\%12\relax}%
\providecommand \@@startlink[1]{}%
\providecommand \@@endlink[0]{}%
\providecommand \url  [0]{\begingroup\@sanitize@url \@url }%
\providecommand \@url [1]{\endgroup\@href {#1}{\urlprefix }}%
\providecommand \urlprefix  [0]{URL }%
\providecommand \Eprint [0]{\href }%
\providecommand \doibase [0]{http://dx.doi.org/}%
\providecommand \selectlanguage [0]{\@gobble}%
\providecommand \bibinfo  [0]{\@secondoftwo}%
\providecommand \bibfield  [0]{\@secondoftwo}%
\providecommand \translation [1]{[#1]}%
\providecommand \BibitemOpen [0]{}%
\providecommand \bibitemStop [0]{}%
\providecommand \bibitemNoStop [0]{.\EOS\space}%
\providecommand \EOS [0]{\spacefactor3000\relax}%
\providecommand \BibitemShut  [1]{\csname bibitem#1\endcsname}%
\let\auto@bib@innerbib\@empty
\bibitem [{\citenamefont {Ratinac}\ \emph {et~al.}(2010)\citenamefont
  {Ratinac}, \citenamefont {Yang}, \citenamefont {Ringer},\ and\ \citenamefont
  {Braet}}]{Ratinac}%
  \BibitemOpen
  \bibfield  {author} {\bibinfo {author} {\bibfnamefont {K.~R.}\ \bibnamefont
  {Ratinac}}, \bibinfo {author} {\bibfnamefont {W.}~\bibnamefont {Yang}},
  \bibinfo {author} {\bibfnamefont {S.~P.}\ \bibnamefont {Ringer}}, \ and\
  \bibinfo {author} {\bibfnamefont {F.}~\bibnamefont {Braet}},\ }\href
  {\doibase 10.1021/es902659d} {\bibfield  {journal} {\bibinfo  {journal}
  {Environmental Science and Technology}\ }\textbf {\bibinfo {volume} {44}},\
  \bibinfo {pages} {1167} (\bibinfo {year} {2010})}\BibitemShut {NoStop}%
\bibitem [{\citenamefont {Schedin}\ \emph {et~al.}(2007)\citenamefont
  {Schedin}, \citenamefont {Geim}, \citenamefont {Morozov}, \citenamefont
  {Hill}, \citenamefont {Blake}, \citenamefont {Katsnelson},\ and\
  \citenamefont {Novoselov}}]{Schedin2007}%
  \BibitemOpen
  \bibfield  {author} {\bibinfo {author} {\bibfnamefont {F.}~\bibnamefont
  {Schedin}}, \bibinfo {author} {\bibfnamefont {A.~K.}\ \bibnamefont {Geim}},
  \bibinfo {author} {\bibfnamefont {S.~V.}\ \bibnamefont {Morozov}}, \bibinfo
  {author} {\bibfnamefont {E.~W.}\ \bibnamefont {Hill}}, \bibinfo {author}
  {\bibfnamefont {P.}~\bibnamefont {Blake}}, \bibinfo {author} {\bibfnamefont
  {M.~I.}\ \bibnamefont {Katsnelson}}, \ and\ \bibinfo {author} {\bibfnamefont
  {K.~S.}\ \bibnamefont {Novoselov}},\ }\href {\doibase 10.1038/nmat1967}
  {\bibfield  {journal} {\bibinfo  {journal} {Nature Materials}\ }\textbf
  {\bibinfo {volume} {6}},\ \bibinfo {pages} {652} (\bibinfo {year}
  {2007})}\BibitemShut {NoStop}%
\bibitem [{\citenamefont {Lafkioti}\ \emph {et~al.}(2010)\citenamefont
  {Lafkioti}, \citenamefont {Krauss}, \citenamefont {Lohmann}, \citenamefont
  {Zschieschang}, \citenamefont {Klauk}, \citenamefont {Klitzing},\ and\
  \citenamefont {Smet}}]{Lafkioti}%
  \BibitemOpen
  \bibfield  {author} {\bibinfo {author} {\bibfnamefont {M.}~\bibnamefont
  {Lafkioti}}, \bibinfo {author} {\bibfnamefont {B.}~\bibnamefont {Krauss}},
  \bibinfo {author} {\bibfnamefont {T.}~\bibnamefont {Lohmann}}, \bibinfo
  {author} {\bibfnamefont {U.}~\bibnamefont {Zschieschang}}, \bibinfo {author}
  {\bibfnamefont {H.}~\bibnamefont {Klauk}}, \bibinfo {author} {\bibfnamefont
  {K.~v.}\ \bibnamefont {Klitzing}}, \ and\ \bibinfo {author} {\bibfnamefont
  {J.~H.}\ \bibnamefont {Smet}},\ }\href {\doibase 10.1021/nl903162a}
  {\bibfield  {journal} {\bibinfo  {journal} {Nano Letters}\ }\textbf {\bibinfo
  {volume} {10}},\ \bibinfo {pages} {1149} (\bibinfo {year}
  {2010})}\BibitemShut {NoStop}%
\bibitem [{\citenamefont {Wang}\ \emph {et~al.}(2010)\citenamefont {Wang},
  \citenamefont {Wu}, \citenamefont {Cong}, \citenamefont {Shang},\ and\
  \citenamefont {Yu}}]{Wang2010}%
  \BibitemOpen
  \bibfield  {author} {\bibinfo {author} {\bibfnamefont {H.}~\bibnamefont
  {Wang}}, \bibinfo {author} {\bibfnamefont {Y.}~\bibnamefont {Wu}}, \bibinfo
  {author} {\bibfnamefont {C.}~\bibnamefont {Cong}}, \bibinfo {author}
  {\bibfnamefont {J.}~\bibnamefont {Shang}}, \ and\ \bibinfo {author}
  {\bibfnamefont {T.}~\bibnamefont {Yu}},\ }\href {\doibase 10.1021/nn101950n}
  {\bibfield  {journal} {\bibinfo  {journal} {ACS Nano}\ }\textbf {\bibinfo
  {volume} {4}},\ \bibinfo {pages} {7221} (\bibinfo {year} {2010})}\BibitemShut
  {NoStop}%
\bibitem [{\citenamefont {Sabri}\ \emph {et~al.}(2009)\citenamefont {Sabri},
  \citenamefont {Levesque}, \citenamefont {Aguirre}, \citenamefont
  {Guillemette}, \citenamefont {Martel},\ and\ \citenamefont
  {Szkopek}}]{Sabri2009}%
  \BibitemOpen
  \bibfield  {author} {\bibinfo {author} {\bibfnamefont {S.~S.}\ \bibnamefont
  {Sabri}}, \bibinfo {author} {\bibfnamefont {P.~L.}\ \bibnamefont {Levesque}},
  \bibinfo {author} {\bibfnamefont {C.~M.}\ \bibnamefont {Aguirre}}, \bibinfo
  {author} {\bibfnamefont {J.}~\bibnamefont {Guillemette}}, \bibinfo {author}
  {\bibfnamefont {R.}~\bibnamefont {Martel}}, \ and\ \bibinfo {author}
  {\bibfnamefont {T.}~\bibnamefont {Szkopek}},\ }\href {\doibase
  DOI:10.1063/1.3273396} {\bibfield  {journal} {\bibinfo  {journal} {Appl.
  Phys. Lett.}\ }\textbf {\bibinfo {volume} {95}},\ \bibinfo {pages} {242104}
  (\bibinfo {year} {2009})}\BibitemShut {NoStop}%
\bibitem [{\citenamefont {Levesque}\ \emph {et~al.}(2011)\citenamefont
  {Levesque}, \citenamefont {Sabri}, \citenamefont {Aguirre}, \citenamefont
  {Guillemette}, \citenamefont {Siaj}, \citenamefont {Desjardins},
  \citenamefont {Szkopek},\ and\ \citenamefont {Martel}}]{Levesque}%
  \BibitemOpen
  \bibfield  {author} {\bibinfo {author} {\bibfnamefont {P.~L.}\ \bibnamefont
  {Levesque}}, \bibinfo {author} {\bibfnamefont {S.~S.}\ \bibnamefont {Sabri}},
  \bibinfo {author} {\bibfnamefont {C.~M.}\ \bibnamefont {Aguirre}}, \bibinfo
  {author} {\bibfnamefont {J.}~\bibnamefont {Guillemette}}, \bibinfo {author}
  {\bibfnamefont {M.}~\bibnamefont {Siaj}}, \bibinfo {author} {\bibfnamefont
  {P.}~\bibnamefont {Desjardins}}, \bibinfo {author} {\bibfnamefont
  {T.}~\bibnamefont {Szkopek}}, \ and\ \bibinfo {author} {\bibfnamefont
  {R.}~\bibnamefont {Martel}},\ }\href {\doibase 10.1021/nl103015w} {\bibfield
  {journal} {\bibinfo  {journal} {Nano Letters}\ }\textbf {\bibinfo {volume}
  {11}},\ \bibinfo {pages} {132} (\bibinfo {year} {2011})}\BibitemShut
  {NoStop}%
\bibitem [{\citenamefont {Sidorov}\ \emph {et~al.}(2011)\citenamefont
  {Sidorov}, \citenamefont {Sherehiy}, \citenamefont {Jayasinghe},
  \citenamefont {Stallard}, \citenamefont {Benjamin}, \citenamefont {Yu},
  \citenamefont {Liu}, \citenamefont {Wu}, \citenamefont {Cao}, \citenamefont
  {Chen}, \citenamefont {Jiang}, ,\ and\ \citenamefont
  {Sumanasekera}}]{Sidorov2011}%
  \BibitemOpen
  \bibfield  {author} {\bibinfo {author} {\bibfnamefont {A.~N.}\ \bibnamefont
  {Sidorov}}, \bibinfo {author} {\bibfnamefont {A.}~\bibnamefont {Sherehiy}},
  \bibinfo {author} {\bibfnamefont {R.}~\bibnamefont {Jayasinghe}}, \bibinfo
  {author} {\bibfnamefont {R.}~\bibnamefont {Stallard}}, \bibinfo {author}
  {\bibfnamefont {D.~K.}\ \bibnamefont {Benjamin}}, \bibinfo {author}
  {\bibfnamefont {Q.}~\bibnamefont {Yu}}, \bibinfo {author} {\bibfnamefont
  {Z.}~\bibnamefont {Liu}}, \bibinfo {author} {\bibfnamefont {W.}~\bibnamefont
  {Wu}}, \bibinfo {author} {\bibfnamefont {H.}~\bibnamefont {Cao}}, \bibinfo
  {author} {\bibfnamefont {Y.~P.}\ \bibnamefont {Chen}}, \bibinfo {author}
  {\bibfnamefont {Z.}~\bibnamefont {Jiang}}, , \ and\ \bibinfo {author}
  {\bibfnamefont {G.~U.}\ \bibnamefont {Sumanasekera}},\ }\href {\doibase
  DOI:10.1063/1.3609858} {\bibfield  {journal} {\bibinfo  {journal} {Appl.
  Phys. Lett.}\ }\textbf {\bibinfo {volume} {99}},\ \bibinfo {pages} {013115}
  (\bibinfo {year} {2011})}\BibitemShut {NoStop}%
\bibitem [{\citenamefont {Moser}\ \emph {et~al.}(2008)\citenamefont {Moser},
  \citenamefont {Verdaguer}, \citenamefont {Jim�nez}, \citenamefont
  {Barreiro},\ and\ \citenamefont {Bachtold}}]{Moser}%
  \BibitemOpen
  \bibfield  {author} {\bibinfo {author} {\bibfnamefont {J.}~\bibnamefont
  {Moser}}, \bibinfo {author} {\bibfnamefont {A.}~\bibnamefont {Verdaguer}},
  \bibinfo {author} {\bibfnamefont {D.}~\bibnamefont {Jim�nez}}, \bibinfo
  {author} {\bibfnamefont {A.}~\bibnamefont {Barreiro}}, \ and\ \bibinfo
  {author} {\bibfnamefont {A.}~\bibnamefont {Bachtold}},\ }\href {\doibase
  DOI:10.1063/1.2898501} {\bibfield  {journal} {\bibinfo  {journal} {Appl.
  Phys. Lett.}\ }\textbf {\bibinfo {volume} {92}},\ \bibinfo {pages} {123507}
  (\bibinfo {year} {2008})}\BibitemShut {NoStop}%
\bibitem [{\citenamefont {Wehling}, \citenamefont {Lichtenstein},\ and\
  \citenamefont {Katsnelson}(2008)}]{Wehling}%
  \BibitemOpen
  \bibfield  {author} {\bibinfo {author} {\bibfnamefont {T.~O.}\ \bibnamefont
  {Wehling}}, \bibinfo {author} {\bibfnamefont {A.~I.}\ \bibnamefont
  {Lichtenstein}}, \ and\ \bibinfo {author} {\bibfnamefont {M.~I.}\
  \bibnamefont {Katsnelson}},\ }\href {\doibase DOI:10.1063/1.3033202}
  {\bibfield  {journal} {\bibinfo  {journal} {Appl. Phys. Lett.}\ }\textbf
  {\bibinfo {volume} {93}},\ \bibinfo {pages} {202110} (\bibinfo {year}
  {2008})}\BibitemShut {NoStop}%
\bibitem [{\citenamefont {Romero}\ \emph {et~al.}(2008)\citenamefont {Romero},
  \citenamefont {Shen}, \citenamefont {Joshi}, \citenamefont {Gutierrez},
  \citenamefont {Tadigadapa}, \citenamefont {Sofo},\ and\ \citenamefont
  {Eklund}}]{Romero2008}%
  \BibitemOpen
  \bibfield  {author} {\bibinfo {author} {\bibfnamefont {H.~E.}\ \bibnamefont
  {Romero}}, \bibinfo {author} {\bibfnamefont {N.}~\bibnamefont {Shen}},
  \bibinfo {author} {\bibfnamefont {P.}~\bibnamefont {Joshi}}, \bibinfo
  {author} {\bibfnamefont {H.~R.}\ \bibnamefont {Gutierrez}}, \bibinfo {author}
  {\bibfnamefont {S.~A.}\ \bibnamefont {Tadigadapa}}, \bibinfo {author}
  {\bibfnamefont {J.~O.}\ \bibnamefont {Sofo}}, \ and\ \bibinfo {author}
  {\bibfnamefont {P.~C.}\ \bibnamefont {Eklund}},\ }\href {\doibase
  10.1021/nn800354m} {\bibfield  {journal} {\bibinfo  {journal} {ACS Nano}\
  }\textbf {\bibinfo {volume} {2}},\ \bibinfo {pages} {2037} (\bibinfo {year}
  {2008})}\BibitemShut {NoStop}%
\bibitem [{\citenamefont {Liao}\ \emph {et~al.}(2010)\citenamefont {Liao},
  \citenamefont {Han}, \citenamefont {Zhou},\ and\ \citenamefont {Yu}}]{Liao}%
  \BibitemOpen
  \bibfield  {author} {\bibinfo {author} {\bibfnamefont {Z.-M.}\ \bibnamefont
  {Liao}}, \bibinfo {author} {\bibfnamefont {B.-H.}\ \bibnamefont {Han}},
  \bibinfo {author} {\bibfnamefont {Y.-B.}\ \bibnamefont {Zhou}}, \ and\
  \bibinfo {author} {\bibfnamefont {D.-P.}\ \bibnamefont {Yu}},\ }\href
  {\doibase DOI:10.1063/1.3460798} {\bibfield  {journal} {\bibinfo  {journal}
  {Appl. Phys. Lett.}\ }\textbf {\bibinfo {volume} {133}},\ \bibinfo {pages}
  {044703} (\bibinfo {year} {2010})}\BibitemShut {NoStop}%
\bibitem [{\citenamefont {Shin}, \citenamefont {Seo},\ and\ \citenamefont
  {Cho}(2011)}]{Shin}%
  \BibitemOpen
  \bibfield  {author} {\bibinfo {author} {\bibfnamefont {W.~C.}\ \bibnamefont
  {Shin}}, \bibinfo {author} {\bibfnamefont {S.}~\bibnamefont {Seo}}, \ and\
  \bibinfo {author} {\bibfnamefont {B.~J.}\ \bibnamefont {Cho}},\ }\href
  {\doibase DOI:10.1063/1.3578396} {\bibfield  {journal} {\bibinfo  {journal}
  {Appl. Phys. Lett.}\ }\textbf {\bibinfo {volume} {98}},\ \bibinfo {pages}
  {153505} (\bibinfo {year} {2011})}\BibitemShut {NoStop}%
\bibitem [{\citenamefont {Lee}\ \emph {et~al.}(2011)\citenamefont {Lee},
  \citenamefont {Kang}, \citenamefont {Jung}, \citenamefont {Kim},
  \citenamefont {Hwang}, \citenamefont {Chung}, \citenamefont {Seo},
  \citenamefont {Choi},\ and\ \citenamefont {Lee}}]{Lee2011}%
  \BibitemOpen
  \bibfield  {author} {\bibinfo {author} {\bibfnamefont {Y.~G.}\ \bibnamefont
  {Lee}}, \bibinfo {author} {\bibfnamefont {C.~G.}\ \bibnamefont {Kang}},
  \bibinfo {author} {\bibfnamefont {U.~J.}\ \bibnamefont {Jung}}, \bibinfo
  {author} {\bibfnamefont {J.~J.}\ \bibnamefont {Kim}}, \bibinfo {author}
  {\bibfnamefont {H.~J.}\ \bibnamefont {Hwang}}, \bibinfo {author}
  {\bibfnamefont {H.-J.}\ \bibnamefont {Chung}}, \bibinfo {author}
  {\bibfnamefont {S.}~\bibnamefont {Seo}}, \bibinfo {author} {\bibfnamefont
  {R.}~\bibnamefont {Choi}}, \ and\ \bibinfo {author} {\bibfnamefont {B.~H.}\
  \bibnamefont {Lee}},\ }\href {\doibase DOI:10.1063/1.3588033} {\bibfield
  {journal} {\bibinfo  {journal} {Appl. Phys. Lett.}\ }\textbf {\bibinfo
  {volume} {98}},\ \bibinfo {pages} {183508} (\bibinfo {year}
  {2011})}\BibitemShut {NoStop}%
\bibitem [{\citenamefont {Chakrapani}\ \emph {et~al.}(2007)\citenamefont
  {Chakrapani}, \citenamefont {Angus}, \citenamefont {Anderson}, \citenamefont
  {Wolter}, \citenamefont {Stoner},\ and\ \citenamefont
  {Sumanasekera}}]{Chakrapani}%
  \BibitemOpen
  \bibfield  {author} {\bibinfo {author} {\bibfnamefont {V.}~\bibnamefont
  {Chakrapani}}, \bibinfo {author} {\bibfnamefont {J.~C.}\ \bibnamefont
  {Angus}}, \bibinfo {author} {\bibfnamefont {A.~B.}\ \bibnamefont {Anderson}},
  \bibinfo {author} {\bibfnamefont {S.~D.}\ \bibnamefont {Wolter}}, \bibinfo
  {author} {\bibfnamefont {B.~R.}\ \bibnamefont {Stoner}}, \ and\ \bibinfo
  {author} {\bibfnamefont {G.~U.}\ \bibnamefont {Sumanasekera}},\ }\href
  {\doibase 10.1126/science.1148841} {\bibfield  {journal} {\bibinfo  {journal}
  {Science}\ }\textbf {\bibinfo {volume} {318}},\ \bibinfo {pages} {1424}
  (\bibinfo {year} {2007})}\BibitemShut {NoStop}%
\bibitem [{\citenamefont {Aguirre}\ \emph {et~al.}(2009)\citenamefont
  {Aguirre}, \citenamefont {Levesque}, \citenamefont {Paillet}, \citenamefont
  {Lapointe}, \citenamefont {St-Antoine}, \citenamefont {Desjardins},\ and\
  \citenamefont {Martel}}]{Aguirre}%
  \BibitemOpen
  \bibfield  {author} {\bibinfo {author} {\bibfnamefont {C.~M.}\ \bibnamefont
  {Aguirre}}, \bibinfo {author} {\bibfnamefont {P.~L.}\ \bibnamefont
  {Levesque}}, \bibinfo {author} {\bibfnamefont {M.}~\bibnamefont {Paillet}},
  \bibinfo {author} {\bibfnamefont {F.}~\bibnamefont {Lapointe}}, \bibinfo
  {author} {\bibfnamefont {B.~C.}\ \bibnamefont {St-Antoine}}, \bibinfo
  {author} {\bibfnamefont {P.}~\bibnamefont {Desjardins}}, \ and\ \bibinfo
  {author} {\bibfnamefont {R.}~\bibnamefont {Martel}},\ }\href {\doibase
  10.1002/adma.200900550} {\bibfield  {journal} {\bibinfo  {journal} {Adv.
  Mater.}\ }\textbf {\bibinfo {volume} {21}},\ \bibinfo {pages} {3087}
  (\bibinfo {year} {2009})}\BibitemShut {NoStop}%
\bibitem [{\citenamefont {Fu}\ \emph {et~al.}(0)\citenamefont {Fu},
  \citenamefont {Nef}, \citenamefont {Knopfmacher}, \citenamefont {Tarasov},
  \citenamefont {Weiss}, \citenamefont {Calame},\ and\ \citenamefont
  {Sch\"{o}nenberger}}]{Fu}%
  \BibitemOpen
  \bibfield  {author} {\bibinfo {author} {\bibfnamefont {W.}~\bibnamefont
  {Fu}}, \bibinfo {author} {\bibfnamefont {C.}~\bibnamefont {Nef}}, \bibinfo
  {author} {\bibfnamefont {O.}~\bibnamefont {Knopfmacher}}, \bibinfo {author}
  {\bibfnamefont {A.}~\bibnamefont {Tarasov}}, \bibinfo {author} {\bibfnamefont
  {M.}~\bibnamefont {Weiss}}, \bibinfo {author} {\bibfnamefont
  {M.}~\bibnamefont {Calame}}, \ and\ \bibinfo {author} {\bibfnamefont
  {C.}~\bibnamefont {Sch\"{o}nenberger}},\ }\href {\doibase 10.1021/nl201332c}
  {\bibfield  {journal} {\bibinfo  {journal} {Nano Letters}\ }\textbf {\bibinfo
  {volume} {0}} (\bibinfo {year} {0}),\ 10.1021/nl201332c}\BibitemShut
  {NoStop}%
\bibitem [{\citenamefont {Ang}\ \emph {et~al.}(2008)\citenamefont {Ang},
  \citenamefont {Chen}, \citenamefont {Wee},\ and\ \citenamefont
  {Loh}}]{Ang2008}%
  \BibitemOpen
  \bibfield  {author} {\bibinfo {author} {\bibfnamefont {P.~K.}\ \bibnamefont
  {Ang}}, \bibinfo {author} {\bibfnamefont {W.}~\bibnamefont {Chen}}, \bibinfo
  {author} {\bibfnamefont {A.~T.~S.}\ \bibnamefont {Wee}}, \ and\ \bibinfo
  {author} {\bibfnamefont {K.~P.}\ \bibnamefont {Loh}},\ }\href {\doibase
  10.1021/ja805090z} {\bibfield  {journal} {\bibinfo  {journal}
  {J.Am.Chem.Soc.}\ }\textbf {\bibinfo {volume} {130}},\ \bibinfo {pages}
  {14392} (\bibinfo {year} {2008})}\BibitemShut {NoStop}%
\bibitem [{\citenamefont {Ohno}\ \emph {et~al.}(2009)\citenamefont {Ohno},
  \citenamefont {Maehashi}, \citenamefont {Yamashiro},\ and\ \citenamefont
  {Matsumoto}}]{Ohno}%
  \BibitemOpen
  \bibfield  {author} {\bibinfo {author} {\bibfnamefont {Y.}~\bibnamefont
  {Ohno}}, \bibinfo {author} {\bibfnamefont {K.}~\bibnamefont {Maehashi}},
  \bibinfo {author} {\bibfnamefont {Y.}~\bibnamefont {Yamashiro}}, \ and\
  \bibinfo {author} {\bibfnamefont {K.}~\bibnamefont {Matsumoto}},\ }\href
  {\doibase 10.1021/nl901596m} {\bibfield  {journal} {\bibinfo  {journal} {Nano
  Letters}\ }\textbf {\bibinfo {volume} {9}},\ \bibinfo {pages} {3318}
  (\bibinfo {year} {2009})}\BibitemShut {NoStop}%
\bibitem [{\citenamefont {Heller}\ \emph {et~al.}(2010)\citenamefont {Heller},
  \citenamefont {Chatoor}, \citenamefont {M\"{a}nnik}, \citenamefont
  {Zevenbergen}, \citenamefont {Dekker},\ and\ \citenamefont {Lemay}}]{Heller}%
  \BibitemOpen
  \bibfield  {author} {\bibinfo {author} {\bibfnamefont {I.}~\bibnamefont
  {Heller}}, \bibinfo {author} {\bibfnamefont {S.}~\bibnamefont {Chatoor}},
  \bibinfo {author} {\bibfnamefont {J.}~\bibnamefont {M\"{a}nnik}}, \bibinfo
  {author} {\bibfnamefont {M.~A.~G.}\ \bibnamefont {Zevenbergen}}, \bibinfo
  {author} {\bibfnamefont {C.}~\bibnamefont {Dekker}}, \ and\ \bibinfo {author}
  {\bibfnamefont {S.~G.}\ \bibnamefont {Lemay}},\ }\href {\doibase
  10.1021/ja104850n} {\bibfield  {journal} {\bibinfo  {journal}
  {J.Am.Chem.Soc.}\ }\textbf {\bibinfo {volume} {132}},\ \bibinfo {pages}
  {17149} (\bibinfo {year} {2010})}\BibitemShut {NoStop}%
\bibitem [{\citenamefont {Veenhuis}\ \emph {et~al.}(2009)\citenamefont
  {Veenhuis}, \citenamefont {van~der Wouden}, \citenamefont {van
  Nieuwkasteele}, \citenamefont {van~den Berg},\ and\ \citenamefont
  {Eijkel}}]{Veenhuis}%
  \BibitemOpen
  \bibfield  {author} {\bibinfo {author} {\bibfnamefont {R.~B.~H.}\
  \bibnamefont {Veenhuis}}, \bibinfo {author} {\bibfnamefont {E.~J.}\
  \bibnamefont {van~der Wouden}}, \bibinfo {author} {\bibfnamefont {J.~W.}\
  \bibnamefont {van Nieuwkasteele}}, \bibinfo {author} {\bibfnamefont
  {A.}~\bibnamefont {van~den Berg}}, \ and\ \bibinfo {author} {\bibfnamefont
  {J.~C.~T.}\ \bibnamefont {Eijkel}},\ }\href {\doibase 10.1039/B913384D}
  {\bibfield  {journal} {\bibinfo  {journal} {Lab Chip}\ }\textbf {\bibinfo
  {volume} {9}},\ \bibinfo {pages} {3472} (\bibinfo {year} {2009})}\BibitemShut
  {NoStop}%
\bibitem [{\citenamefont {Kim}(2010)}]{Kim}%
  \BibitemOpen
  \bibfield  {author} {\bibinfo {author} {\bibfnamefont {S.~H.}\ \bibnamefont
  {Kim}},\ }in\ \href@noop {} {\emph {\bibinfo {booktitle} {Advanced
  Tribology}}},\ \bibinfo {editor} {edited by\ \bibinfo {editor} {\bibfnamefont
  {J.}~\bibnamefont {Luo}}, \bibinfo {editor} {\bibfnamefont {Y.}~\bibnamefont
  {Meng}}, \bibinfo {editor} {\bibfnamefont {T.}~\bibnamefont {Shao}}, \ and\
  \bibinfo {editor} {\bibfnamefont {Q.}~\bibnamefont {Zhao}}}\ (\bibinfo
  {publisher} {Springer Berlin Heidelberg},\ \bibinfo {year} {2010})\ pp.\
  \bibinfo {pages} {584--585}\BibitemShut {NoStop}%
\bibitem [{\citenamefont {Huard}\ \emph {et~al.}(2008)\citenamefont {Huard},
  \citenamefont {Stander}, \citenamefont {Sulpizio},\ and\ \citenamefont
  {Goldhaber-Gordon}}]{Huard}%
  \BibitemOpen
  \bibfield  {author} {\bibinfo {author} {\bibfnamefont {B.}~\bibnamefont
  {Huard}}, \bibinfo {author} {\bibfnamefont {N.}~\bibnamefont {Stander}},
  \bibinfo {author} {\bibfnamefont {J.~A.}\ \bibnamefont {Sulpizio}}, \ and\
  \bibinfo {author} {\bibfnamefont {D.}~\bibnamefont {Goldhaber-Gordon}},\
  }\href {\doibase 10.1103/PhysRevB.78.121402} {\bibfield  {journal} {\bibinfo
  {journal} {Phys. Rev. B}\ }\textbf {\bibinfo {volume} {78}},\ \bibinfo
  {pages} {121402} (\bibinfo {year} {2008})}\BibitemShut {NoStop}%
\bibitem [{\citenamefont {Teoh}\ \emph {et~al.}(2011)\citenamefont {Teoh},
  \citenamefont {Tao}, \citenamefont {Tok}, \citenamefont {Ho},\ and\
  \citenamefont {Sow}}]{Teoh}%
  \BibitemOpen
  \bibfield  {author} {\bibinfo {author} {\bibfnamefont {H.~F.}\ \bibnamefont
  {Teoh}}, \bibinfo {author} {\bibfnamefont {Y.}~\bibnamefont {Tao}}, \bibinfo
  {author} {\bibfnamefont {E.~S.}\ \bibnamefont {Tok}}, \bibinfo {author}
  {\bibfnamefont {G.~W.}\ \bibnamefont {Ho}}, \ and\ \bibinfo {author}
  {\bibfnamefont {C.~H.}\ \bibnamefont {Sow}},\ }\href {\doibase
  DOI:10.1063/1.3580762} {\bibfield  {journal} {\bibinfo  {journal} {Appl.
  Phys. Lett.}\ }\textbf {\bibinfo {volume} {98}},\ \bibinfo {pages} {173105}
  (\bibinfo {year} {2011})}\BibitemShut {NoStop}%
\bibitem [{\citenamefont {Hansen}(2007)}]{Hansen}%
  \BibitemOpen
  \bibfield  {author} {\bibinfo {author} {\bibfnamefont {C.~M.}\ \bibnamefont
  {Hansen}},\ }\href@noop {} {\emph {\bibinfo {title} {Hansen Solubility
  Parameters: A User's Handbook}}}\ (\bibinfo  {publisher} {CRC Press},\
  \bibinfo {year} {2007})\BibitemShut {NoStop}%
\end{thebibliography}
\providecommand{\noopsort}[1]{}\providecommand{\singleletter}[1]{#1}%

\end{document}